\documentclass[%
 aip,
rsi,%
 amsmath,amssymb,
 reprint,%
]{revtex4-1}
\usepackage{textcomp}
\usepackage{graphicx}
\usepackage{dcolumn}
\usepackage{bm}
\usepackage{siunitx}

\begin{document}

\preprint{AIP/123-QED}

\title[Stockholm, \today]{A stable wavelength-tunable triggered source of single photons and cascaded photon pairs at the telecom C-band}

\author{Katharina~D.~Zeuner}
\author{Matthias~Paul}
\author{Thomas~Lettner}
\affiliation{Department of Applied Physics, Royal Institute of Technology, Albanova University Centre, Roslagstullsbacken 21, 106 91 Stockholm, Sweden}%
\author{Carl~Reuterski\"old~Hedlund}
\affiliation{Department of Electronics, Royal Institute of Technology, Electrum 229, 164 40 Kista, Sweden}
\author{Lucas~Schweickert}
\affiliation{Department of Applied Physics, Royal Institute of Technology, Albanova University Centre, Roslagstullsbacken 21, 106 91 Stockholm, Sweden}%
\author{Stephan~Steinhauer}
\author{Lily~Yang}
\affiliation{Department of Applied Physics, Royal Institute of Technology, Albanova University Centre, Roslagstullsbacken 21, 106 91 Stockholm, Sweden}%
\author{Julien~Zichi}
\affiliation{Department of Applied Physics, Royal Institute of Technology, Albanova University Centre, Roslagstullsbacken 21, 106 91 Stockholm, Sweden}%
\author{Mattias~Hammar}
\affiliation{Department of Electronics, Royal Institute of Technology, Electrum 229, 164 40 Kista, Sweden}
\author{Klaus~D.~J\"ons}
\email[Corresponding author: ]{klausj@kth.se}
\affiliation{Department of Applied Physics, Royal Institute of Technology, Albanova University Centre, Roslagstullsbacken 21, 106 91 Stockholm, Sweden}%
\author{Val~Zwiller}
\affiliation{Department of Applied Physics, Royal Institute of Technology, Albanova University Centre, Roslagstullsbacken 21, 106 91 Stockholm, Sweden}%

\date{\today}

\begin{abstract}

The implementation of fiber-based long--range quantum communication requires tunable sources of single photons at the telecom C--band. Stable and easy--to--implement wavelength--tunability of individual sources is crucial to (i) bring remote sources into resonance, to (ii) define a wavelength standard and to (iii) ensure scalability to operate a quantum repeater.
So far the most promising sources for true, telecom single photons are semiconductor quantum dots, due to their ability to deterministically and reliably emit single and entangled photons. However, the required wavelength--tunability is hard to attain. Here, we show a stable wavelength--tunable quantum light source by integrating strain--released InAs quantum dots on piezoelectric substrates. We present triggered single--photon emission at $1.55\,\si{\micro\meter}$ with a multiphoton emission probability as low as $0.097$, as well as photon pair emission from the radiative biexciton--exciton cascade. We achieve a tuning range of $0.25\,\si{\nano\meter}$ which will allow to spectrally overlap remote quantum dots or tuning distant quantum dots into resonance with quantum memories. This opens up realistic avenues for the implementation of photonic quantum information processing applications at telecom wavelengths.

\end{abstract}

\pacs{Valid PACS appear here}
\maketitle
%
%
Creating single photons on--demand is one of the essential requirements to connect remote nodes of a quantum network using quantum repeaters. In the past years, semiconductor quantum dots (QDs) have proven to be excellent sources of on--demand single,\cite{He2013,Senellart2017,Schweickert2017} 
indistinguishable photons, \cite{Santori,Somaschi2016,Ding2016} and entangled photon pairs.\cite{Muller2014} 
Those photons will play the crucial role of flying qubits in quantum communication and quantum information processing. QDs are scalable, via the expanding field of nanofabrication straight-forward to integrate and be electrically driven with high excitation rates.\cite{Hargart2013} 
Furthermore, QDs possess a tailorable emission wavelength based on a wide range of available semiconductor materials and growth methods. In order to enable long--range optical communication, emission wavelengths within the telecommunication C--band ($1530-1565\,\si{\nano\meter}$) are required, due to the absorption minimum in the glass fiber material at these wavelengths. These emission wavelengths are long compared to the well--studied emission of typical InAs/GaAs QDs and have, previously, only been realized in the InAs/InP material system. \cite{Miyazawa2005,Birowosuto2012, Benyoucef2013} However, the latter material system is brittle, distributed Bragg mirrors are hard to realize, and the optical quality of the QDs is lagging behind compared to QDs on GaAs substrates. Very recently, the first C--band QDs in the InGaAs/GaAs material system were fabricated \cite{Paul2017} by carefully engineering the interface between substrate and QD layer, thus, providing a surface with reduced lattice mismatch. The application of such a metamorphic buffer layer (MMBL) \cite{Ledentsov2003, Semenova2008} allows emission of single and entangled photon pairs at wavelengths of $1.55\,\si{\micro\meter}$ \cite{Olbrich2017}. However, based on the stochastic growth process of self-assembled QDs, there is a distribution in size and shape within a QD ensemble, as well as fluctuations in the QD's electrical environment, leading to a distribution of the emission wavelength. Hence, spectrally overlapping photons from different QDs requires either the inefficient process of searching for nearly identical QDs or applying emission energy tuning mechanisms. To facilitate this, several tuning mechanisms for semiconductor QDs have been studied in the recent past, i.e. tuning via temperature, lateral and vertical electric \cite{Kowalik2005,Gerardot2011,Marcet2010} 
or magnetic \cite{Bayer2002} fields, as well as strain. \cite{Seidl2006,Jons2011}
Tuning the QD emission wavelength via temperature, however, increases the interaction of charge carriers with phonons, thereby degrading the single-photon purity and introducing additional spectral broadening. Using electric fields possesses the disadvantage of separating electrons and holes from each other, leading to decreased emission rates. Finally, tuning via magnetic fields requires the integration of large magnets into the setup, making it unfeasible for an application. In contrast, strain tuning offers a cheap and easily--integrable method that allows the controll of the optical properteis of the QDs. Recently, complex piezo--tuning devices have been realized that allow the simultaneous tuning of QD emission energy and the FSS, thus, enabling the generation of energy--tunable entangled photon pairs. \cite{Trotta2016,Chen2016} 
To the best of our knowledge, none of the described tuning techniques have been reported to realize a QD device showing tunable single-photon emission at $1.55\,\si{\micro\meter}$. In particular, piezoelectric strain tuning has so far not been employed yet for QDs emitting at telecom wavelengths, due to the challenging InP material system rendering integration impossible. Here, we report on piezoelectric strain tuning of the emission wavelength from a mechanically thinned sample with InGaAs/GaAs QDs emitting at the telecom C-band. 
%
%
%
%
The sample is fabricated in a commercial AIX-200 horizontal flow reactor at a pressure of 100\,mbar on exactly oriented (100), Si-doped GaAs substrates. As precursors, we used TMGa, TMIn, TMAl, and AsH$_3$. In a first step, we deposited a distributed Bragg reflector (DBR) consisting of 20 alternating pairs of AlAs/GaAs with a nominal thickness of \SI{112}{\nano\metre} and \SI{95}{\nano\metre}, respectively, followed by an InGaAs layer of $1150\,\si{\nano\meter}$ and an additional $280\,\si{\nano\meter}$ thick InGaAs capping layer. This structure forms a 3$\lambda$ cavity between DBR and the semiconductor-air interface. We optimized the cavity for a wavelength of \SI{1550}{\nano\metre}. The single layer of InAs QDs is deposited on top of the metamorphic InGaAs layer. This position lies in an anti-node of the standing electric wave and, thus, leads to an increase of the coupling between excitonic states and resonator. For the metamorphic InGaAs layer, we increased the In flux during the growth time, i.e., the Indium concentration in the buffer increases with thickness, decreasing the InAs/InGaAs lattice mismatch. More details on the sample growth are described in Ref~\onlinecite{Paul2017}.
%
%
%
%
\begin{figure}[htb]
\includegraphics[width=\columnwidth]{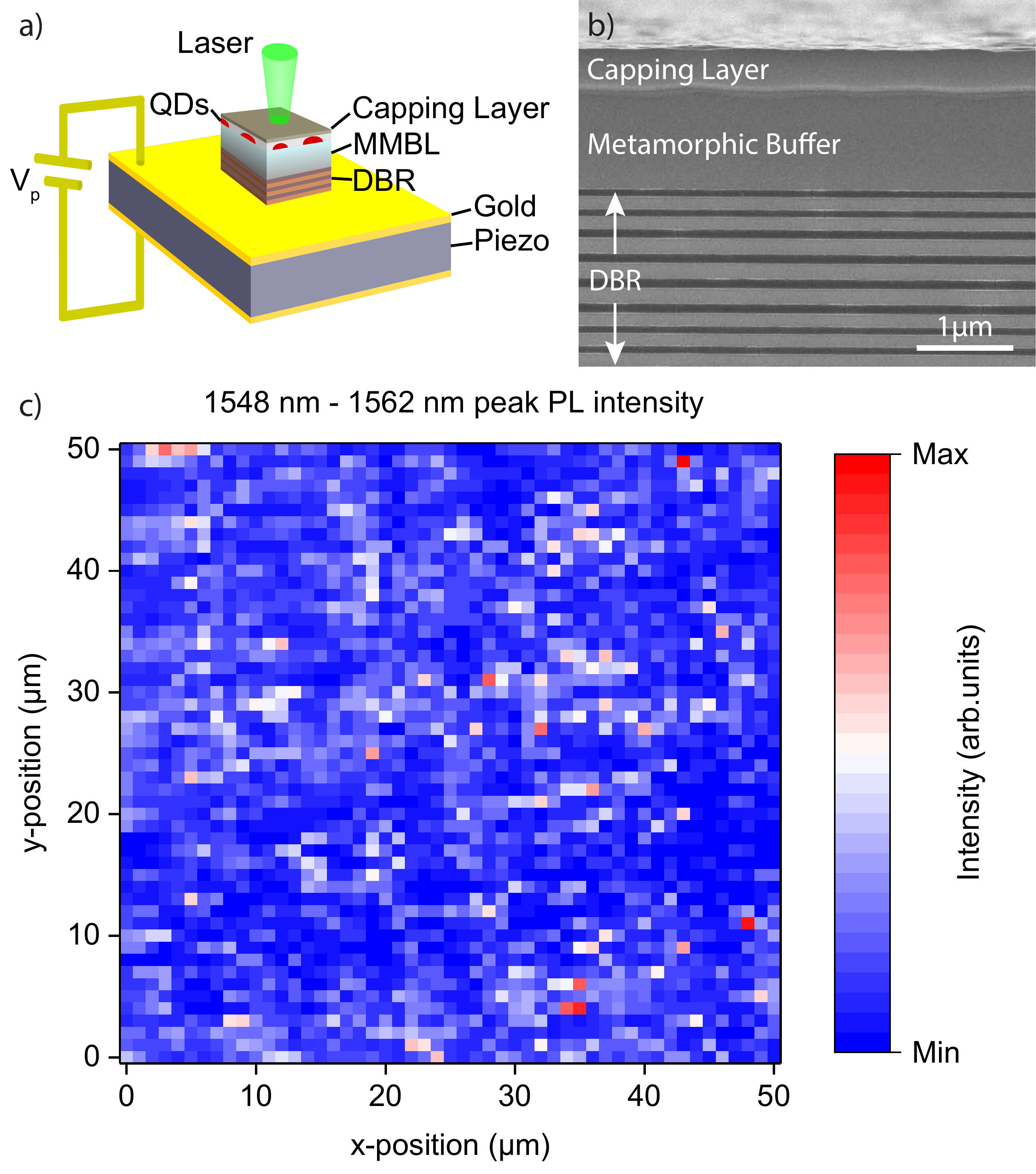}
\caption{\label{fig:sample} (a) Sample schematic: the mechanically thinned sample glued to a piezoelectric element which is connected to a PCB via compression bonding. For simplicity the remaining substrate is not depicted. (b) SEM-image of the a cleaved facet of the non-thinned sample. From bottom to top: top DBR layers, metamorphic buffer layer, QDs and capping layer. (c) Photoluminescence map of the sample in a $50\times50\,\si{\micro\meter}$ range. Displayed is the PL peak intensity in a wavelength range from $1548-1562\,\si{\nano\meter}$.}
\end{figure}
For stable and reversible energy tuning of our quantum dot emission we integrate the sample on a gold coated  PMN-PT piezoelectric actuator (TRS technologies, thickness of $200\,\si{\micro\meter}$, $\langle 001 \rangle$ orientation). In order to increase the tuning range the quantum dot sample is mechanically thinned using diamond--based abrasive films. The mechanical lapping of the sample allows for the integration of samples with arbitrary epitaxial layer structures (including DBRs), in contrast to chemical etching methods using a sacrificial layer.~\cite{Zander2009,Rastelli2012} The sample surface is glued to the end of a brass rod with removable wax to apply uniform pressure during the lapping process. The sample thickness is monitored using a plunger dial. After the desired thickness of approximately $50\,\si{\micro\meter}$ is reached, the sample is released from the polishing rod in an acetone bath. Afterwards the sample is carefully sieved from the bath using soft membrane tissue. We developed a reliable transfer method exploiting electrostatic forces to pick and place the thinned sample on the prepared piezo. By using a bendable soft tip to which the sample attaches during transfer, we are able to handle samples with thicknesses down to $15\,\si{\micro\meter}$. As a glue we use cryogenic epoxy (Stycast ES--2--20, two component resin) to achieve a rigid connection between the sample and the piezo actuator as well as good thermal contact. We attach the piezoelectric actuator only in one of its corners to the copper sample mount with conductive silver paste in order to guarantee unrestricted contraction and expansion of the piezo and, thus, transfer the maximum stress to the sample. A schematic of the sample design is shown in figure \ref{fig:sample} a) illustrating from bottom to top: gold--coated piezo (for electrical contacting), DBR layers, MMBL with QDs on top and the capping layer. Between the top and bottom of the piezoelectric element a voltage can be applied to strain the sample.
%
%
%
%
The sample is cooled to about $10\,\si{\kelvin}$ in a closed--cycle cryostat. In our confocal microscopy setup the sample is excited trough a microscope objective ($0.85\,\si{NA}$, $100\times\,\si{magnification}$). A stack of three piezoelectric positioners is used in order to precisely position the QD under investigation underneath the microscope objective and furthermore allows us to scan the sample during excitation. We perform above--band excitation of the QD by either using a narrowband continuous wave (cw) laser at $795\,\si{\nano\meter}$ or a tunable pulsed laser with a repetition rate of $80\,\si{\mega\hertz}$ and a pulse duration of $2\,\si{\pico\second}$. The QD photons are coupled into a single-mode fiber and is then sent either to a spectrometer that is equipped with an InGaAs array for spectral analysis or to superconducting single--photon detectors (SSPDs) optimized for telecom wavelengths to perform correlation measurements. For the spectral filtering during the correlation measurements we use tunable fiber--based filters with a $3\,\si{\decibel}$--bandwidth of $0.8\,\si{\nano\meter}$.
%
%
%
%
We take scanning electron microscope (SEM) images of the cleaved facet of the non-thinned sample to investigate layer thicknesses as presented in Fig. \ref{fig:sample} b). From bottom to top, the uppermost layers of the DBR, the metamorphic buffer, the QD layer and the capping layer are visible. For the thickness of the metamorphic buffer we measure $1.15\,\si{\micro\meter}$ and the capping layer thickness is $250-300\,\si{\nano\meter}$. We observe undulations in the thickness of the MMBL, which is consistent with the findings in Ref.~\onlinecite{Paul2017}. These undulations are also observed for the capping layer thickness. For further characterization, we record photoluminescence (PL) maps by scanning the sample in  $1\,\si{\micro\meter}$ steps in an area of $50\times50\,\si{\micro\meter}$ during above--band cw excitation. At each position we take a spectrum with an integration time of $2\,\si{\second}$ in a spectral range from $1548\,\si{\nano\meter}-1562\,\si{\nano\meter}$. The result is presented in Fig. \ref{fig:sample} c), where the peak intensity is plotted over the position of the piezo actuators with a read--out accuracy of $500\,\si{nm}$. The PL map allows us to asses not only the density of optically active QDs, but also distribution on the sample. We infer a QD density on the order of $10^7\,\si{\centi\meter}^{-2}$. Assuming a spot diameter on the sample of $1\,\si{\micro\meter}$, we find on average only one QD in the excitation spot, which is ideal for single QD spectroscopy. 
%
%
%
%

\begin{figure}[htb]
\includegraphics[width=\columnwidth]{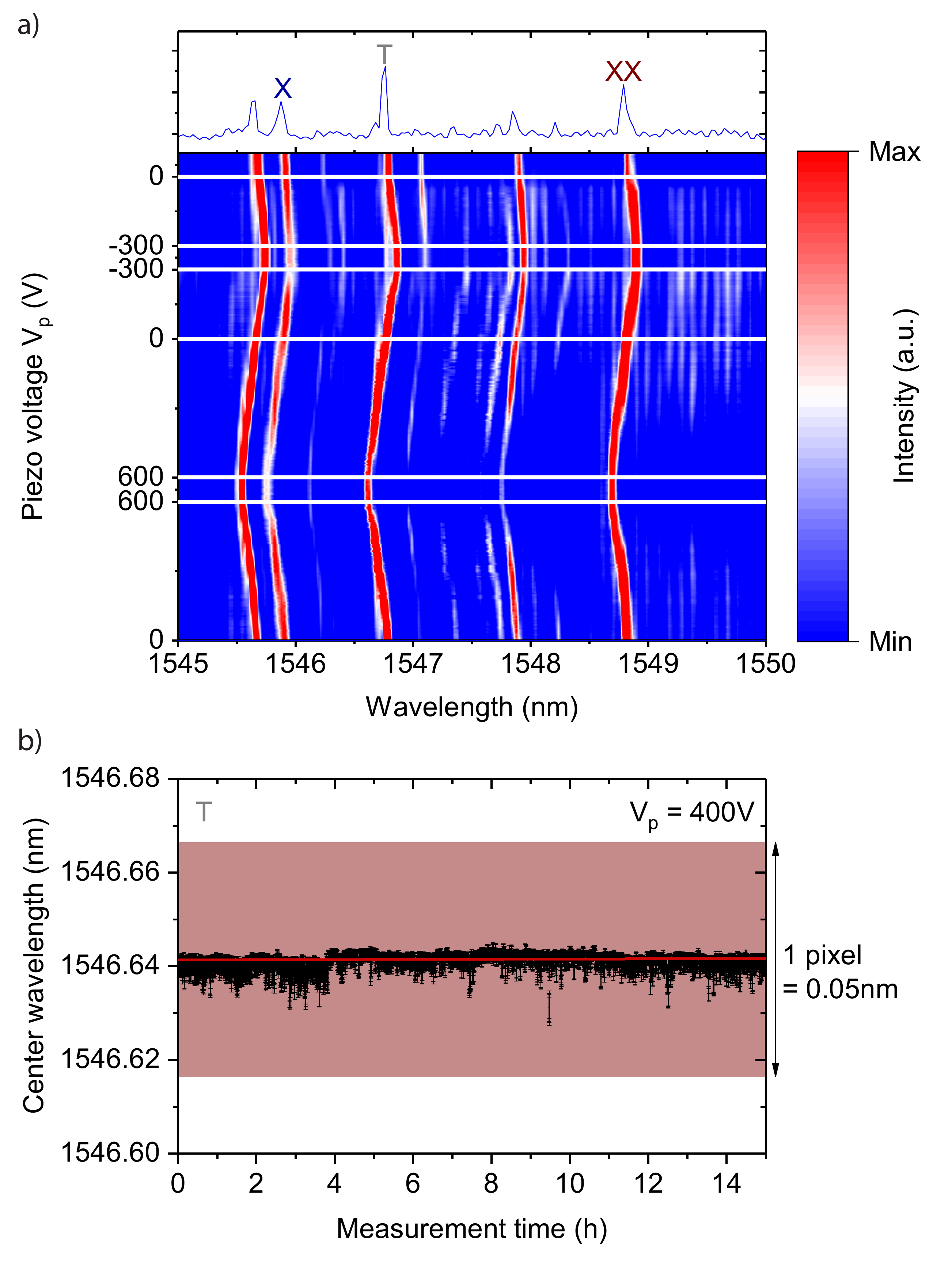}
\caption{\label{fig:tuning} (a) Top image: PL spectrum of QD 1 at $0\,\si{\volt}$, denoted are exciton (X), trion (T) and biexciton (XX) state, the intensity is plotted in arbitrary units. Bottom image: QD emission as a function of applied voltage to the piezo in an overall range of $900\,\si{\volt}$. (b) Center wavelength of the T line for an applied voltage of $+400\,\si{\volt}$ plotted over a measurement duration of 15\,h. The red area spans over the bandwidth of a single pixel of our CCD camera. Marked with a red line is a fit to the data.}
\end{figure}
In Figure \ref{fig:tuning} a) we present the spectrum of QD 1 with emission at the telecom C--band. The spectrum consists of an exciton (X), a trion (T) and a biexciton (XX) transition. The three different s-shell states were identified via power and polarization dependent measurements. Below the spectrum we present a tuning map, displaying the emission wavelength as a function of the applied voltage to the piezoelectric actuator. Positive (negative) voltage corresponds to in--plane compression (expansion) of the positively poled piezo actuator.
We operate the piezo in a voltage range from $+600\,\si{\volt}$ to $-300\,\si{\volt}$ to ensure repeatability and reproducibility of the strain tuning. The voltage was changed in $1\,\si{\volt}$ steps and for every step we take a spectrum  with an integration time of $10\,\si{\second}$. Starting point is the emission of the sample at $0\,\si{\volt}$, then the applied voltage is increased to $+600\,\si{\volt}$ and then held there for $1000\,\si{\second}$, while continuously recording spectra. Following this, the applied voltage is reduced to $-300\,\si{\volt}$, and then again ramped to $0\,\si{\volt}$, with stops for $1000\,\si{\second}$ at $-300\,\si{\volt}$ and $0\,\si{\volt}$, again marked by vertical lines. Tuning towards higher positive voltages corresponds to a blue-shift of the emission, whereas tuning towards lower voltages leads to a red-shift of the emission wavelengths. We extract the tuning range of the emission wavelength from fits to the spectra. We find a tuning range of $0.25\,\si{\nano\meter}$, which corresponds to $0.28\,\si{\pico\meter\volt}^{-1}$, that allows extremely precise tuning of quantum dots emission. Furthermore, the observed wavelength shift is reversible and follows the same linear behavior for all lines under investigation. While a broad tuning range is desired in order to be able to reach a target wavelength with a large number of QDs, the longterm stability of the emission once tuned to the target wavelength is of equal importance for reliable operation. An investigation of the emission wavelength of the trion--line over a measurement period of $15\,\si{\hour}$ for an applied voltage of $400\,\si{\volt}$ is shown in Figure \ref{fig:tuning} b). The individual center wavelength of each spectrum is obtained via fits to the data, as the change in wavelength over time is clearly below the resolution limit of the spectrometer. We mark the spectral width of one pixel on the CCD, which corresponds to $25\,\si{\micro\eV}$, to further illustrate the longterm stability of our tunable source. A linear fit to the data is shown in figure \ref{fig:tuning} a), we find a slope of zero within the fit error. This order of stability enables a longterm overlap of QD photons with respect to a specific resonance.\cite{Zopf2017} 
%
%
%
%

\begin{figure}[htb]
\includegraphics[width=\columnwidth]{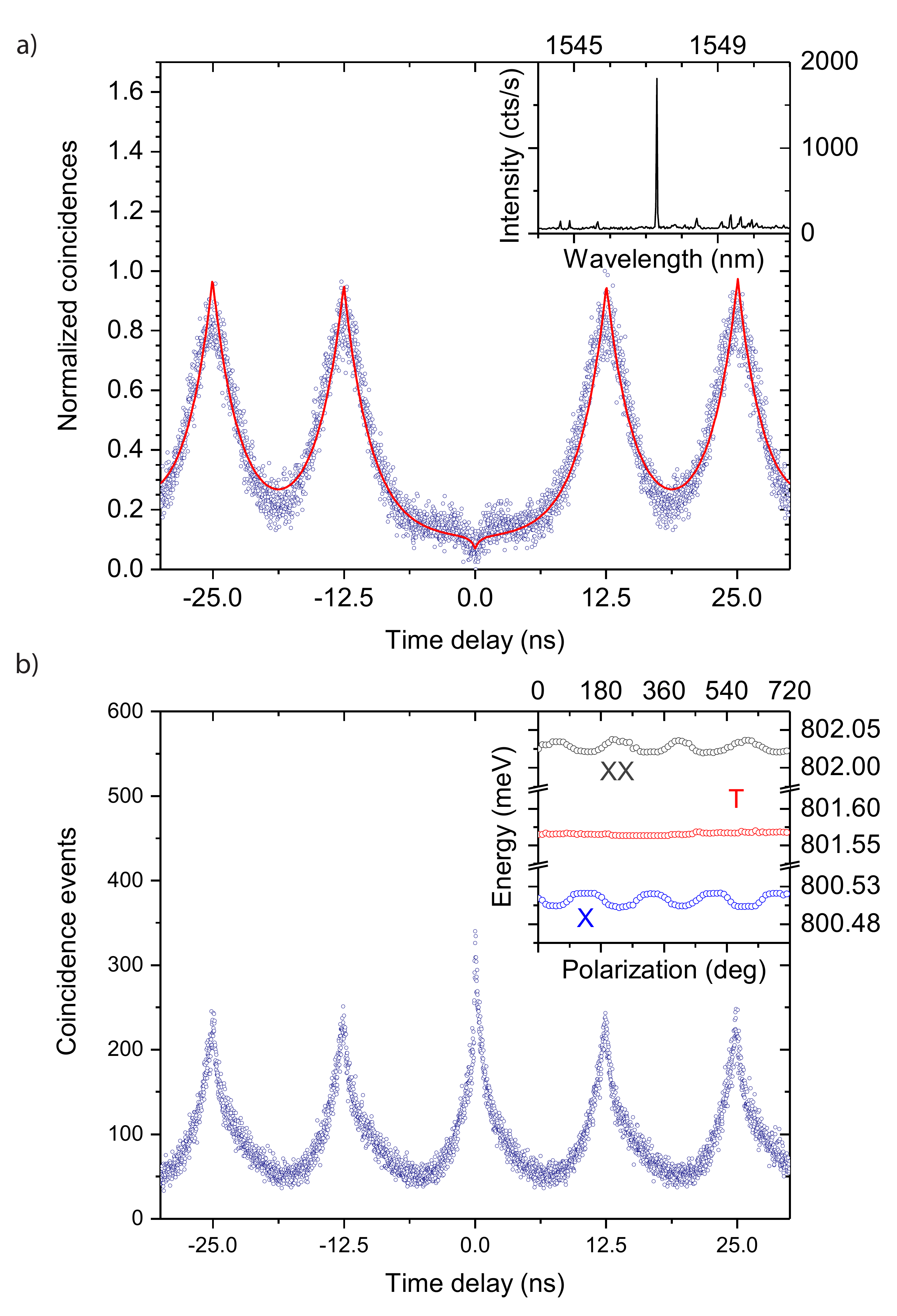}
\caption{\label{fig:correlation} (a) Second--order auto--correlation function measured under pulsed excitation of QD 2, spectrum in inset. (b) Cross--correlation function of biexciton (XX) and exciton (X) of QD 1. Inset: result of polarization--dependent measurement of QD 1, shown for X, T and XX.}
\end{figure}

Lastly, we investigate photon statistics properties of the telecom C--band quantum dots emission. A second--order auto--correlation measurement is performed to analyze the multiphoton emission probability. For this, the QD under investigation (QD 2) is excited with $2\,\si{\pico\second}$ pulses at $780\,\si{\nano\meter}$ and the photons of the emission line (Figure \ref{fig:correlation} a), inset) are filtered with a tunable fiber--based filter and then sent onto a $50:50$ beam splitter. A start--stop measurement is carried out with two SSPDs, the correlation histogram is recorded and displayed in Figure \ref{fig:correlation} a). Strong suppression of the peak at zero time delay is apparent which corresponds to a low probability of multiphoton emission. The contribution of multiphoton events is calculated via a fit to the data according to Ref.~\onlinecite{Nakajima2012} and gives a value of $0.097\pm0.045$. The fit takes the population dynamics of the quantum two--level system into account and is able to reproduce the dip observed at zero time delay. 
Furthermore, we perform a cross--correlation measurement of the X and XX of QD~1 under pulsed excitation. The emission of QD~1 is split with a $50:50$ beam splitter and then a fiber--based filter in each arm is set to either the X ($1546.9\,\si{\nano\meter}$) or the XX wavelength ($1549.9\,\si{\nano\meter}$). Subsequently, a start--stop measurement between the XX and X photons is performed. The existence of a XX--X--cascade is clearly visible in Figure \ref{fig:correlation} b). Once the XX state is decayed, the probability of the emission of an X photon increases, leading to the observed bunching peak for zero time delay, thus, confirming the existence of a radiative cascade. 
Cascaded emission of photon pairs via the radiative cascade results in polarization entangled photon pairs.\cite{Benson2000}~However, for the QD under investigation exhibits a finestructure--splitting $34.8\,\si{\micro eV}$, as shown in the inset of Figure \ref{fig:correlation} b), leading to a fast time--evolving entangled state.\cite{Ward2014}~
%
%
%
%
\\
In conclusion, we have demonstrated precise emission energy tuning of single and cascaded photons at $1550\,\si{\nano\meter}$ with high precision, using a piezoelectric substrate. We achieved a tuning range of $0.25\,\si{\nano\meter}$, which is 
enough to tune QDs in a spectral distance of $0.5\,\si{\nano\meter}$ into resonance and, thus, demonstrates the potential of strain tuning for telecom QDs. 
Furthermore, we show that the targeted wavelength after tuning is stable, which allows to perform longterm experiments with this type of device. 
The demonstrated wavelength tunability will be of crucial importance to spectrally overlap remote QDs for entanglement swapping operations and tuning QDs into resonance with quantum memories. The authors thank Raffaele Gallo for contributions to the sample growth. The Quantum Nano Photonics Group at KTH acknowledges the continuous support by the company APE on their picoEmerald system. K.D.Z. gratefully acknowledges funding by the Dr. Isolde Dietrich Foundation. L.Y. ackowledges funging by VINNOVA under grant No. 2015-04928. K.D.J. acknowledges funding from the MARIE SK\L ODOWSKA-CURIE  Individual  Fellowship under REA grant agreement No. 661416 (SiPhoN). V.Z. acknowledges funding by the European Research Council under the grant agreement No. 307687 (NaQuOp) and the Swedish Research Council under grant agreement 638-2013-7152. Financial support was also provided by the Linnaeus Center in Advanced Optics and Photonics (Adopt). 
\bibliography{telecom_strain_paper}

\end{document}